\documentclass[twocolumn,showpacs,preprintnumbers,amsmath,amssymb]{revtex4}

\usepackage{epsfig}
\usepackage{graphicx}
\usepackage{dcolumn}

\newcommand{\beq}{\begin{equation}}
\newcommand{\eeq}{\end{equation}}
\newcommand{\beqa}{\begin{eqnarray}}
\newcommand{\eeqa}{\end{eqnarray}}
\newcommand{\lam}{\lambda}

\newcommand{\rh}{\rho}

\newcommand{\si}{\sigma}

\newcommand{\om}{\omega}

\newcommand{\la}{\langle}
\newcommand{\ra}{\rangle}

\def\jmo#1{{ J.\ Mod.\ Opt.} {\bf#1}}
\def\jpb#1{{ J.\ Phys.\ B} {\bf#1}}

\def\pr#1{{ Phys.\ Rev. } {\bf#1}}
\def\pra#1{{ Phys.\ Rev. A\/} {\bf#1}}

\def\prl#1{{ Phys.\ Rev.\ Lett.} {\bf#1}}
\def\qic#1{{ Quant.\ Inf.\ Comp.} {\bf#1}}
\def\sci#1{{ Science} {\bf#1}}

\begin{document}

\title{Coherent State Control of Non-Interacting Quantum Entanglement}


\author{Muhammed Y\"ona\c{c} and J.H. Eberly}
\affiliation{Rochester Theory Center, and Department of Physics and
Astronomy, University of Rochester, New York 14627, USA }

\begin{abstract}
We exploit a novel approximation scheme to obtain a new and compact formula for the parameters underlying coherent-state control of the evolution of a pair of entangled two-level systems. It is appropriate for long times and for relatively strong external quantum control via coherent state irradiation. We take account of both discrete-state and continuous-variable degrees of freedom. The formula predicts the relative heights of entanglement revivals and their timing and duration.  \end{abstract}

\maketitle

\section{Introduction}

Almost 75 years after its specific introduction by Schr\"odinger \cite{Schr} a general understanding of entanglement within quantum theory is still being sought in the sense that measures of it are still being developed. Only two-particle entanglement is well quantified, and even in that case the dimension of the state space that can be taken into account is extremely small unless the joint state is pure. The behavior of few-state entangled quantum systems takes a wide variety of forms in different contexts, such as cavity QED, circuit QED, spontaneous parametric down conversion, photonic lattices and ion traps, involving atoms, ions, photons, superconducting circuits, quantum dots, and spins. In any of these cases, the controlled manipulation of entanglement with external agents is one approach to a form of quantum control, and it remains an open challenge.

\begin{figure}[!b]
\begin{center}
\includegraphics[scale=0.4]{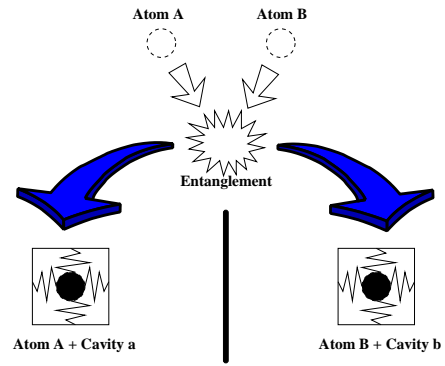}\end{center}
\caption[A diagram illustrating our system.]{\label{setup}  (Color online)  A diagram illustrating our system.  The atoms $A$ and $B$ are placed in their respective cavities $a$ and $b$ after being entangled. Only $A$-$a$ and $B$-$b$ interactions take place thereafter, but non-interacting $A$-$B$ entanglement changes in time.}\end{figure}

We will examine preservation rather than generation of entanglement, particularly the evolution of two remote qubits that are entangled and being ``stored" for later use, ideally isolated from environmental influence, but subject to localized control interactions. The qubits have the character of a primitive quantum memory unit. Our approach differs further from that of most previous studies by focusing on the effects of (i) strong control fields, (ii) that are treated quantum mechanically but are not themselves initially entangled, and (iii) which act for times significantly longer than Rabi oscillation time scales. We will work under the condition that the members of the entangled pair are truly remote, i.e., do not interact with each other directly, or through third parties even indirectly. We will take the control mechanism as a continuous variable (CV) interaction, which is treated using the Jaynes-Cummings (JC) model Hamiltonian \cite{Jaynes-Cummings}. A sketch of a cavity QED context for this study is shown in Fig. \ref{setup}.

There are previous theoretical discussions that also use the JC interaction for related considerations arising in the interaction of a qubit pair with CV fields. These are typically interested in questions of entanglement generation in the qubits by transfer from initially weakly squeezed states of CV cavity fields. An early study by Kim, et al. \cite{Kim-etalJOM02}, considered the action of a two-mode squeezed field that is injected onto two qubits held in two separate high-Q cavities. From the two-qubit reduced density matrix, obtained by tracing over the fields, they calculated subsequent atom-atom entanglement, and found that it is maximized when the squeezed field state is pure. Their results include one of the first examples of the time-dependent entanglement birth and death effect (see overview \cite{Yu-EberlySci09}). The question whether and to what extent the entanglement of weak squeezing could be transferred to the initially unentangled qubits was answered positively.

Later studies \cite{Krause-Cirac, Paternostro-etalPRL04, Paternostro-etalPRA04} showed that the field entanglement in the JC cavity could in principle be employed for transfer into networks or to subsequent pairs of atoms transitting the cavities. This work was extended \cite{Kim-etalPRL06} to address an example of the ``storage" problem, and transfer of entanglement from discrete qubits to an entangled coherent state was also examined by Zhou and Wang \citep{Zhou-Wang}. We comment below on relations to our results.  The evolution of entanglement in a qubit-field system, where the qubit and the field start from mixed states was examined by Rendell and Rajagopal \cite{Rendell-Rajagopal}. Because of the lack of an entanglement measure for $2\times\infty$ systems, their calculation of entanglement for the entire system was limited to a lower bound for the concurrence.

We first quickly review entanglement dynamics in the JC context in order to establish notation for control parameters, time scales, etc., and to allow direct interpretation of a new approximate formula - see relation (\ref{C vs T}) below - for entanglement control parameters. We restrict attention to the dynamics of entanglement between two sites holding qubits $A$ and $B$. We suppose that there is quantum memory present in the qubits that is of interest, i.e., that the two atoms have been entangled in some way before being inserted into their respective cavities. Field modes $a$ and $b$ are available at the $A$-$B$ sites that can be used for interacting with the atoms (i.e., for ``controlling" them and their entanglement externally). This scenario is suggested in Fig. \ref{setup}.

\section{Hamiltonian Equations}

The Jaynes-Cummings \citep{Jaynes-Cummings} Hamiltonian (with $\hbar = 1$) is given by:
\beqa
H_{\rm tot} &=& \frac{\om_0}{2}\sigma_z^A  +
(g_A^* a^{\dagger}\sigma_{-}^A + g_A\sigma_{+}^A a)  + \om a^{\dagger}a \nonumber\\
 && + \frac{\om_0}{2}\sigma_z^B + (g_B^*b^{\dagger}\sigma_{-}^B +
 g_B\sigma_{+}^B b) + \om b^{\dagger}b,
\eeqa
where $\om_0$ is the transition frequency between the two levels of the atoms, $g$ is the constant of coupling between the atoms and the fields and $\om$ is the angular frequency of the single-mode field. The usual Pauli matrices $\sigma_z^{A,B},\ \sigma_+^{A,B}$ and $\sigma_-^{A,B}$ describe the atoms, while $a^{\dagger},\ a$ and $b^{\dagger},\ b$ are the raising and lowering operators for the fields. The difference between $g$ and $g^*$ in both cases is only a phase, but there may be a substantial difference between $g_A$ and $g_B$ if the cavities are different.

\begin{figure}[!b]
\centerline{\includegraphics[scale=0.7]{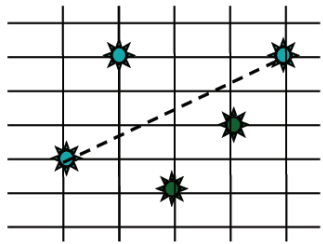}}
  \caption[Sketch indicating non-interacting qubits in a
quantum storage network.]{(Color online) Sketch indicating non-interacting qubits in a
quantum storage network.  Quantum memory in the form of two-qubit entanglement between any two relatively remote sites, indicated here by a dashed line, can be altered by fields local to the sites.} \label{f.network-square}
\end{figure}

The Hamiltonian does not support any interaction between atom $A$ and atom $B$ or between mode $a$ and mode $b$.  Physical realization of our scenario does not appear out of the question, as the Jaynes-Cummings model has been realized in the laboratory in several well-known ways \citep{Harochegroup, Winelandgroup, Kimblegroup, Rempegroup}.

To illustrate the approach to be followed when the fields are taken as highly excited (nearly classical) coherent states,  we start with a simpler example, in which the field modes are initially in their vacuum states: $|0_a\rangle\otimes |0_b\rangle$, and the two atoms are in a superposition of Bell states.  An additional simplification, to focus on the most elementary aspects of the entanglement evolution first, will be to take the cavities identical, and assume $g_A = g_B \equiv g$.  For the vacuum case the coupling terms in our Hamiltonian indicate that for creation of a photon in a cavity the atom in that cavity must decay to a lower state.  Since the cavities each contain only one atom, there can be no more than one photon in each at any time. This means that the cavities are also two-state quantum systems, i.e., qubits.

The eigenstates of the JC Hamiltonian are well known.  We will write for either $Aa$ or $Bb$
\beq
\label{JCEvalEqn} H_{\rm JC}|\psi_n^{\pm}\ra = \lambda_n^{\pm}
|\psi_n^{\pm}\ra.
\eeq
We will denote the excited and ground atomic states by $|e\ra$ and $|g\ra$, and denote the cavity modes' photon states by the photon number $n$. Then the JC eigenvalues are given by:
\beq \label{JCEvals} \lambda_n^{\pm} = n\omega +
\frac{1}{2}\Big(\Delta \pm \sqrt{\Delta^2 + G_n^2}\Big), \eeq
where $\Delta = \om_0- \om$ is the detuning and
\beq
G_n = 2g\sqrt{n},
\eeq
is the $n$-photon Rabi frequency.

The JC eigenstates have the following form as superpositions of the bare atom and cavity states:
\beqa
\label{JCstates}
|\psi_0\ra &=& |g;0\ra \\
|\psi_n^+\ra &=& c_n|e; n-1\ra + s_n|g; n\ra\,\,\, (n>0)\\
|\psi_n^-\ra &=& -s_n|e; n-1\ra + |c_n|g; n\ra\,\,\, (n>0).
\eeqa
In these equations we have introduced some convenient abbreviations:
\beq
\label{cands}
c_n \equiv \cos(\theta_n/2) \quad {\rm and}\quad s_n \equiv
\sin(\theta_n/2),
\eeq
where the rotation angle $\theta_n$ can be
identified with the Bloch sphere polar angle and is defined in the
usual way:
\beq \label{thetadef}
\cos\theta_n \equiv
\frac{\Delta}{\sqrt{\Delta^2 + G_n^2}} \quad {\rm and} \quad
\sin\theta_n \equiv \frac{G_n}{\sqrt{\Delta^2 + G_n^2}}.
\eeq

In this preliminary example we will need the true ground state $|\psi_0\ra = |g;0\ra$ and the two dressed states for $n=1$. These three states are closed under the JC Hamiltonian for each site.  In other words,  we use only $n=1$ in the equations above, so the subscript $n$ can mostly be ignored and we will frequently drop it ($\lam_n \to \lam$, $c_n \to c$, etc.).

\section{Measure of Pairwise Entanglement}
\label{concurrence/standard}

The JC dressed states are atom-photon locally-entangled states themselves, and this entanglement has interesting consequences if the cavities are fairly highly excited, as Gea-Banacloche  \cite{Banacloche} and Phoenix and Knight \cite{Phoenix-Knight} originally discussed. Here we are interested in non-local atom-atom entanglement between the two spatially separated sites on a qubit lattice, as suggested in Fig. \ref{f.network-square}.

In the general context of entanglement we note that there is no accepted and practically workable criterion for determining separability of arbitrary four-particle states. Our purposes will be satisfied by working with the two-qubit atomic states obtained from the time-evolving four-qubit state. All familiar measures agree about separability in the two-qubit domain of entanglement. That is, entropy of formation, Schmidt number, tangle, negativity, and concurrence are numerically somewhat different, but in the two-qubit domains of their applicability they are in full agreement when they signal entanglement or lack of entanglement.

We adopt Wootters' concurrence \citep{Wootters} as our measure in this discussion, mainly for its convenient normalization: $1 \ge C \ge 0$, where $C=0$ indicates separability (zero entanglement) and $C=1$ means maximal pure state entanglement, as in a Bell state; and simplicity of calculation:
\beq \label{definationc}
C(\rh) = \max\{0,\sqrt{\lam_1} - \sqrt{\lam_2} - \sqrt{\lam_3} - \sqrt{\lam_4} \},
\eeq
where the quantities $\lam_i$ are the eigenvalues in decreasing order of the auxilliary matrix
\beq \zeta=\rho(\sigma_y\otimes
\sigma_y)\rho^*(\sigma_y\otimes \sigma_y),
\label{concurrence}
\eeq
where $\rh^*$ denotes the complex
conjugation of $\rh$ in the standard basis  and
$\si_y$ is the Pauli matrix expressed in the standard basis.

One finds that in reduction to two-qubit form, by tracing over the two field-mode qubits, the resulting two-qubit mixed state always has the $X$ form \citep{Yu-EberlyQIC07}, where only the diagonal and anti-diagonal elements are not zero:
\beq
\label{mixedstate0}
\rho = \left[
\begin{array}{cccc}
\rho_{11} & 0 & 0 & \rho_{14} \\
0 & \rho_{22} & \rho_{23} & 0 \\
0 & \rho_{23}^* & \rho_{33} & 0 \\
\rho_{14}^* & 0 & 0 & \rho_{44}
\end{array} \right], \\
\eeq
where $\rho_{11}+\rho_{22}+\rho_{33}+\rho_{44} = 1$. The concurrence of this mixed state is easily found to be
\beqa \label{generalC}
C &=& 2\max\{0, |\rho_{23}| - \sqrt{\rho_{11}\rho_{44}}, |\rho_{14}|-\sqrt{\rho_{22}\rho_{33}}\}\nonumber \\
&\equiv& 2\max\{0, Q\},
\eeqa
so it is clear that $Q$, which is the larger of $|\rho_{23}| - \sqrt{\rho_{11}\rho_{44}}$ and $|\rho_{14}|-\sqrt{\rho_{22}\rho_{33}}$, will be an important quantity. For example, $Q(t)$ obeys certain conservation relations (whereas $C$ does not) in some special cases because it can be negative. Furthermore, one can see \cite{Yu-EberlyJMO07} that while $Q \le 0$ implies a separable state ($C = 0$), the slightly stronger condition $Q < 0$ implies both $C = 0$ and also that the state is mixed rather than pure, information not available from $C$.

\section{Partially entangled Bell states}
\label{firstpair}

In the two-site situation under consideration there are, in principle, six clearly distinguishable entanglements. They have been overlooked in the earlier papers mentioned \cite{ Kim-etalJOM02, Krause-Cirac, Paternostro-etalPRL04, Paternostro-etalPRA04, Kim-etalPRL06, Zhou-Wang, Rendell-Rajagopal}, but they all can carry information about the bipartite entanglements that may arise. With an obvious notation we can denote these concurrences as $C^{AB}$, $C^{ab}$, $C^{Aa}$, $C^{Bb}$, $C^{Ab}$, $C^{Ba}$.  Except for $C^{Aa}$ and $C^{Bb}$, they measure remote (non-local) entanglements, and their dynamics have been reported to be restricted by previously unknown invariant combinations of entanglement parameters \cite{Yonac-etal07, Sainz-Bjork, Chan-etal}. For background we summarize results when the initial states are superpositions of the Bell states: $|\Phi^{\pm)}\ra \sim |e_A, e_B\ra \pm |g_A, g_B\ra $, which we write:
\beq
\label{PhiZero1}
|\Phi_{AB}\ra =\cos\alpha|e_A, e_B\ra + \sin\alpha|g_A, g_B\ra.
\eeq
It is easy to see that  $\alpha = \pm\pi/4$  reproduces the two pure Bell states, and other values of $\alpha$ represent an adjustable phase between the entangled atomic states. The optional value of $\alpha$ is significant because it can be understood as a measure of the relative difference in preparation of the two atoms in the two identicial cavities. This can be important from the point of view of experimental relevance, as identical preparation of the atoms would be difficult.

The initial state for the atoms plus cavities is therefore:
\beqa
\label{PZero}
|\Phi(0)\ra & = &
|\Phi_{AB}\ra \otimes |0_a,0_b\ra \\
&=&
(\cos\alpha|e_A, e_B\ra + \sin\alpha|g_A, g_B\ra) \otimes |0_a,0_b\ra. \nonumber \eeqa
In terms of the dressed eigenstates given above (\ref{JCstates}), we can rewrite:
\beqa \label{inverseJCstates}
|e_A, 0_a\ra &=& c|\psi_1^+\ra - s|\psi_1^-\ra\nonumber\\
|g_A, 1_a\ra &=& s|\psi_1^+\ra + c|\psi_1^-\ra \quad {\rm
and}\nonumber \\
|g_A, 0_a\ra &=& |\psi_0\ra.
\eeqa
Thus the initial atom-atom entangled state has the form
\beqa
&&\cos\alpha|e_A,0_a\ra\otimes|e_B,0_b\ra  + \sin\alpha |g_A, 0_a\ra \otimes |g_B,0_b\ra \nonumber\\
&& = \cos\alpha (c|\psi_1^+\ra_A - s|\psi_1^-\ra_A) \otimes  (c|\psi_1^+\ra_B - s|\psi_1^-\ra_B)\nonumber \\
&& {}\quad  + \sin\alpha|\psi^0\ra_A \otimes|\psi^0\ra_B.
\eeqa

Evolution in time is easily followed through the evolution of
the dressed states:
\beq \label{psiEvolution}
|\psi^{\pm}(t)\ra = e^{-i\lambda^{\pm}t}~|\psi^{\pm}(0)\ra. \eeq
Note that since the combination of coefficients in $|\Phi(0)\ra$  uniquely associates $c$ with $|\psi^+\ra$, and $s$ with $|\psi^-\ra$, the time evolution can be transferred to the $c$ and $s$ symbols. We will henceforth consider them carrying the time-evolution exponents. We will use the notation $c_0 \equiv c(0)$ and $s_0 \equiv s(0)$ to refer to their values at $t=0$ (no relation to the $n=0$ subscripts in Eq.~(\ref{cands})), so that,
 \beqa
 c = c(t) &=&c_0 e^{-i\lambda^+t},\nonumber\\
 s = s(t) &=&s_0 e^{-i\lambda^-t},
 \eeqa
where $\lambda^+$ and $\lambda^-$ are obtained by inserting $n=1$ into Eq. (\ref{JCEvals}).  Then we can write (temporarily again indicating explicit time dependences for the $c$'s and $s$'s):
\beqa |\Phi(t)\ra &=& \cos\alpha \Big(c(t) |\psi_1^+\ra_A
- s(t)|\psi_1^-\ra_A \Big) \nonumber \\
& \otimes & \Big(c(t)|\psi_1^+\ra_B - s(t)|\psi_1^-\ra_B\Big) \nonumber \\
&+& \sin\alpha|\psi^0\ra_A \otimes|\psi^0\ra_B,
\eeqa
where $|\psi^{\pm}\ra$ will continue to refer to the JC states at $t=0$.

Now we revert to the bare basis states in preparation for the tracing  needed to calculate $Q^{AB}$ and $C^{AB}$:
\beqa
\label{psifunction}
|\Phi(t)\ra & =& \cos\alpha \Big(c(t)(c_0|e_A,0_a\ra + s_0|g_A,1_a\ra) \nonumber\\
&-& s(t)(-s_0|e_A,0_a + c_0|g_A, 1_a\ra)\Big)
\otimes \Big( c(t)(c_0|e_B,0_b\ra \nonumber \\
&+& s_0|g_B,1_b\ra) - s(t)(-s_0|e_B,0_b + c_0|g_B, 1_b\ra)\Big) \nonumber \\
&+& \sin\alpha|g_A, 0_a\ra \otimes |g_B, 0_b\ra \nonumber \\
&=& \cos\alpha\Big((c(t)c_0 + s(t)s_0)|e_A,0_a\ra \nonumber \\
&+& (c(t)s_0 -s(t)c_0)|g_A, 1_a\ra \Big) \nonumber \\
&\otimes& \Big((c(t)c_0 +s(t)s_0)|e_B,0_b\ra \nonumber \\
&+& (c(t)s_0 -s(t)c_0)|g_B, 1_b\ra \Big) \nonumber\\
&+& \sin\alpha |g_A,0_a\ra \otimes |g_B,0_b\ra. \eeqa

To get the two-qubit mixed state needed for calculation of $Q^{AB}$ the projections that are needed are:
\beqa
\la 0_a, 0_b|\Phi(t)\ra &=& \cos\alpha(cc_0 + ss_0)^2|e_a,e_B\ra + \sin\alpha |g_A, g_B\ra \nonumber \\
\la 1_a,0_b|\Phi(t)\ra &=& \cos\alpha (cs_0 - sc_0)(cc_0 + ss_0) |g_A,e_B\ra \nonumber \\
\la 0_a,1_b|\Phi(t)\ra &=& \cos\alpha (cc_0 + ss_0)(cs_0 - sc_0) |e_A,g_B\ra\nonumber\\
\la 1_a,1_b|\Phi(t)\ra &=&0,
\eeqa
where now $c$ and $s$ have replaced $c(t)$ and $s(t)$ to conserve space.  These results show that the $AB$ mixed state has the $X$ form mentioned in the previous Section:
\beq
\label{mixedstate}
\rho^{AB} = \left[
\begin{array}{cccc}
\rho_{11} & 0 & 0 & \rho_{14} \\
0 & \rho_{22} & 0 & 0 \\
0 & 0 & \rho_{33} & 0 \\
\rho_{14}^* & 0 & 0 & \rho_{44}
\end{array} \right], \\
\eeq
for which the concurrence has the stated form
\beq \label{concurrenceEqn}
C^{AB} = 2 ~max\{0, |\rho_{14}|-\sqrt{\rho_{22}\rho_{33}}\}.
\eeq
and we easily find the following
\beqa
|\rho_{14}| &=& |\sin\alpha \cos\alpha|(c_0^4 + s_0^4 +2c_0^2s_0^2\cos\delta t), \nonumber \\
b = c &=& \cos^2\alpha |cc_0 + ss_0|^2~|cs_0 - sc_0|^2 \nonumber \\
&=& \cos^2\alpha (c_0^4 + s_0^4 +2c_0^2s_0^2\cos\delta t) \nonumber \\
& \times & c_0^2 s_0^2 (2-2\cos\delta t),
\eeqa
where $\delta=\sqrt{\Delta^2 + 4g^2}$.

For simplicity we will evaluate this in the resonance case,
$\theta_n = \pi/2$, where $c_0 = s_0 = 1/\sqrt{2}$ and $\Delta=0$. Then we find
\beq
|\rho_{14}| - \sqrt{\rho_{22}\rho_{33}} = \frac{1}{4}\cos^2\alpha [2+2\cos(gt)][|\tan\alpha| - \sin^2(gt)],
\eeq
from which the expression for concurrence is found to be:
\beq
\label{e.CAB-ESD} C^{AB} = 2\max\{0, Q^{AB} \},
\eeq
where $Q^{AB} =\cos^2\alpha\cos^2(gt)[|\tan\alpha| - \sin^2(gt)]$.

\begin{figure}[!h]
\begin{center}
\includegraphics[scale=0.4]{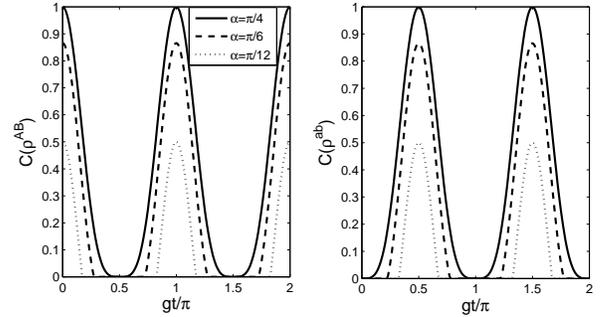}
\caption[]{\label{eeggESD} Time dependences of AB and ab concurrences for three values of the superposition angle $\alpha$. Note that almost all curves show a time interval over which C=0, i.e., during which the underlying state must be separable. The time scale is the vacuum Rabi period. From ref. \cite{Yonac-etal07}.}
\end{center}
\end{figure}

Fig. \ref{eeggESD} shows that for $\alpha \ne \pi/4$ the $C^{AB}$ curves have the ``sudden death" feature \cite{Yu-EberlySci09}. That is, the entanglement non-smoothly becomes zero and stays zero for a finite interval of time. The death and rebirth dynamics of non-interacting entanglement shown is however still governed by the same Rabi time scale that one associates with vacuum Rabi oscillations \cite{JJSM, Agarwal84}.

\section{Open-System Two-Qubit Theory}

\begin{figure}[!t]
\begin{center}
\includegraphics[width=8cm]{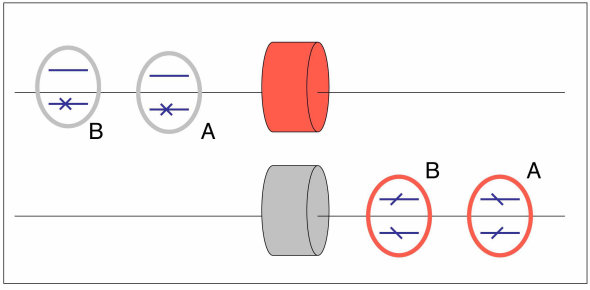}
\caption[]{\label{f.Bellcreation} (Color Online) Top and bottom sketches show a pair of atoms entering and then leaving a prepared resonant cavity, indicating in principle a preparation of the Bell State $\Psi_{AB}$ in advance of insertion into the two cavities that are shown in Fig. \ref{setup}. This is a post-selection approach, as described in the text.}
\end{center}
\end{figure}

Realistic quantum control almost necessarily implies engagement of continuous variables and the interaction of qubits or other systems having a finite number of states with one or more ``large" systems with continuous quantum states. We can use the JC formalism described above to enter this domain by introducing coherent state fields at the $AB$ network sites.  In Fig. \ref{f.Bellcreation} we indicate a post-selection method for experimentally obtaining a Bell State to work with, in principle explaining the entanglement stage left open in Fig. \ref{setup}. That is, the cavity shown in Fig. \ref{f.Bellcreation} is used to prepare atoms to be inserted into the two cavities in Fig. \ref{setup}. It is initially prepared in a single photon state, e.g., by micromaser methods. After the atoms pass through it, each entering one of the final cavities shown in Fig. \ref{setup} to begin their JC interactions, this cavity is monitored for several lifetimes, to ensure that its photon was transferred to one of the atoms $A$ and $B$ (without knowledge which one), and hasn't remained in the cavity. That run of the experiment is cancelled if/when a cavity photon leaks out, and the experiment must be restarted with another pair of ground-state atoms.

We retain almost all of the simplifying approximations made earlier, and add one more by taking the two fields, now modeled as coherent state fields, to have the same average photon number $\bar n$, which could be arranged experimentally by feeding both cavities from the same coherent state laser field via a 50-50 beam splitter. However, one important simplification that we relied on previously must now be discarded. The coherent-state fields have many occupied photon-number states, so the cavities will not be two-state qubits. We assume initial atom entanglement in the form of the pure Bell State prepared, e.g., as in Fig. \ref{f.Bellcreation}:
\beq \label{BellState}
|\Psi_{AB}(0)\rangle = (|eg\rangle+|ge\rangle)/\sqrt{2},
\eeq
and write the field state as the coherent state product
\beq \label{CohState}
|\Psi_{ab}(0)\rangle = |\alpha\rangle\otimes|\alpha\rangle.
\eeq
As a result our initial state for the whole system is,
\beq
|\Psi_{tot}(0)\rangle = |\Psi_{AB}(0)\rangle \otimes |\alpha\rangle \otimes |\alpha\rangle.
\eeq
The coherent states are given by the familiar Fock state expansion
\beq
|\alpha\rangle=\sum_{n=0}^{\infty}A_n |n\rangle = \sum_{n=0}^{\infty} \frac{e^{-|\alpha|^2/2}\alpha^{n}}{\sqrt{n!}}|n\rangle.
\eeq

Now the photonic density matrix is infinite-dimensional, and the joint $AB$ dynamics extremely complicated. Fully numerical analysis is possible, but the insights from analytic results are highly desirable. We have found a key step permitting analytic progress. This is an apparently drastic simplification of the continuous state spaces of the two field modes. We assume that it is satisfactory to replace $|\alpha\rangle$ by $|\bar n\rangle$. This Ansatz is at least weakly supported by the knowledge that photon number in a coherent state is Poisson-distributed and relatively tightly centered around $\bar{n}$ when $\bar n \gg 1$. Thus we represent the initial field state by the single product Fock state $|\bar n\rangle\otimes|\bar n\rangle$. Note that during the JC interaction the photon number in either field mode $a$ or $b$ can then be $\bar n$ or $\bar n + 1$ or $\bar n - 1$.

Under the Ansatz mentioned, by tracing the field mode states, we find that the reduced density matrix for the qubits becomes:
\begin{equation}\label{Xstate}
\rho = \left(
\begin{array}{cccc}
      \rho_{11} & x & x & x \\
      x & \rho_{22} & \rho_{23} & x \\
      x & \rho_{23}* & \rho_{33} & x \\
      x & x & x & \rho_{44} \\
    \end{array}
\right) \to \left(
\begin{array}{cccc}
      \rho_{11} & 0 & 0 & 0 \\
      0 & \rho_{22} & \rho_{23} & 0 \\
      0 & \rho_{23}* & \rho_{33} & 0 \\
      0 & 0 & 0 & \rho_{44} \\
    \end{array} \right),
\end{equation}
where we have used the standard two-qubit  basis [$ee,\ eg,\ ge,\
gg$]. The elements indicated by $x$ are zero because of the
equal-$\bar n$ simplification.  Thus, under the assumptions
mentioned, $\rho$ is again of $X$-type \citep{Yu-EberlyQIC07}).

For this $X$-type $\rho$, concurrence turns into:
\beq
C(\rho) = 2\ max[~ 0,\ |\rho_{23}|-\sqrt{\rho_{11}\rho_{44}}~ ].
\eeq
The coherent fields induce time-dependent change of the elements $\rho_{11}$ and $\rho_{44}$, and their growth and any decline of $\rho_{23}$ will cause entanglement to decrease.

In order to calculate the time evolution of this state under the JC Hamiltonian we need to calculate the time evolution of the the states of the individual sites, i.e., $|e\rangle \otimes |n\rangle$ and $|g\rangle \otimes |n\rangle$ for all $n$.  The time evolution of these states for site $A$ (and similarly for site $B$) is given by,
\beqa
e^{i H_I^A t}|e;n\rangle &=& \cos(gt\sqrt{n+1})|e;n\rangle \nonumber \\
&-& i\sin(gt\sqrt{n+1})|g;n+1\rangle \\
e^{i H_I^A t}|g;n\rangle &=& \cos(gt\sqrt{n})|g;n\rangle \nonumber \\
&-& i\sin(gt\sqrt{n})|e;n-1\rangle.
\eeqa

Using these results, the time evolution of the initial state of the system is found to be,
\beqa
|\Psi_{tot}(t)\rangle &=& e^{i H_I^A t}|\Psi_{tot}(0)\rangle\\
&=&\frac{1}{\sqrt{2}}\sum_{n=0}^{\infty}\sum_{m=0}^{\infty}A_n A_m \Big(K_{mn}\Big),
\eeqa
where
\beqa
K_{mn} &=& -iC_{n+1}S_m|e,e;n,m-1\rangle
+ C_{n+1}S_{m}|e,g;n,m\rangle \nonumber\\
&-&S_{n+1}S_n|g,e;n+1,m\rangle -iS_{n+1}C_n|g,g;n+1,m\rangle \nonumber\\
&-&iS_nC_{m+1}|e,e;n-1,m+1\rangle \nonumber \\
&-&S_nS_{m+1}|e,g;n-1,m+1\rangle\nonumber\\
&+&C_nC_{m+1}|g,e;n,m+1\rangle \nonumber \\
&-&iC_nS_{m+1}|g,g;n,m+1\rangle,
\eeqa
where $C_n=\cos(gt\sqrt{n})$ and $S_n=\sin(gt\sqrt{n})$.  The density matrix for the system is then given by
\beq
\rho_{tot}=|\Psi_{tot}(t)\rangle \langle |\Psi_{tot}(t)|.
\eeq
and the reduced density matrix for the atoms, $\rho_{AB}$, is given by
$\rho_{AB}=Tr_{(n,m)}\rho_{tot}$.

\section{Formulas for Control Parameters}

It is obvious that the quantities denoted $\rho_{23},\ \rho_{11},\ \rho_{44}$ are the control parameters for atom entanglement. Having used the Fock state shortcut to obtain (\ref{Xstate}), we now avoid using it further and introduce the approximation method that leads to analytic formulas for $\rho_{23},\ \rho_{11},\ \rho_{44}$ appropriate to $\rho_{AB}$ for
the coherent state.

One can show that the $\rho_{23}$ term is given by the doubly infinite summation,
\beqa \label{element_z}
z &=& \frac{1}{2}\Big\{\sum_{n,m}A_n^2A_m^2C_nC_{n+1}C_mC_{m+1}\nonumber\\
&-&A_nA_{n-1}A_mA_{m+1}S_nC_{n+1}C_mS_{m+1}\nonumber\\
&+&A_nA_{n-2}A_mA_{m+2}S_nS_{n-1}S_{m+1}S_{m+2}\nonumber\\
&-&A_nA_{n-1}A_mA_{m+1}S_nC_{n-1}S_{m+1}C_{m+2}\Big\},
\eeqa
Similarly the series summations for $\rho_{11}$ and $\rho_{44}$ are:
\beqa \label{element_a}
a &=& \frac{1}{2}\Big\{\sum_{n,m}A_n^2A_m^2C_{n+1}^2S_m^2\nonumber\\
&+&A_nA_{n+1}A_mA_{m-1}S_{n+1}C_{n+1}S_mC_m\nonumber\\
&+&A_n^2A_m^2S_n^2C_{m+1}^2\nonumber\\
&+&A_nA_{n-1}A_mA_{m+1}S_nC_nS_{m+1}C_{m+1}\Big\}
\eeqa
and
\beqa \label{element_d}
d &=& \frac{1}{2}\Big\{\sum_{n,m}A_n^2A_m^2S_{n+1}^2C_m^2\nonumber\\
&+&A_nA_{n+1}A_mA_{m-1}S_{n+1}C_{n+1}S_mC_m\nonumber\\
&+&A_n^2A_m^2C_n^2S_{m+1}^2\nonumber\\
&+&A_nA_{n-1}A_mA_{m+1}S_nC_nS_{m+1}C_{m+1}\Big\}. \eeqa
The infinite extent of these summations of course reflects the fact
that we are dealing with a quantum open system, by having coupled the qubits to an infinite state space.


The sums above cannot be completed, but excellent analytic approximations can be found for coherent states that are even only moderately strong, i.e., $\alpha \ge 10$. We will use the familiar  Stirling formula for $n!$,
\beq
n!=\sqrt{2\pi n} n^n e^{-n},
\eeq
and Euler's formula to approximate the terms in the summations above
by integrals.  We begin by approximating the terms like
$A_nA_{n+1}A_mA_{m-1}$ with $A_n^2 A_m^2$, which introduces an error
of order $1/\bar{n}$ near the Poisson peaks $n \approx m \approx
\bar n$. Then we get for $z$
\beqa \label{new_z}
\rho_{23} &\cong& \frac{1}{2}\Big\{\Big(\sum_{n}A_n^2C_nC_{n+1}\Big)^2\nonumber\\
&-&2\Big(\sum_{n,m}A_n^2A_m^2S_nC_{n+1}C_mS_{m+1}\Big)\nonumber\\
&+&\Big(\sum_{n}A_n^2S_nS_{n+1}\Big)^2\Big\}.
\eeqa
In the same way, Eq.(\ref{element_a}) and Eq.(\ref{element_d}) become,
\beq \label{new_a}
\rho_{11} \cong \Big((\sum_{n}A_nC_n^2\Big)\Big(\sum_n A_n^2S_n^2\Big)+\Big(\sum_{n}A_n^2 S_n C_n\Big)^2,
\eeq
\beqa \label{new_d}
\rho_{44} \cong \Big(\sum_{n}A_nC_n^2\Big)\Big(\sum_n A_n^2S_n^2\Big)+\Big(\sum_{n}A_n^2 S_n C_n\Big)^2.
\eeqa

Note that Eq. (\ref{new_a}) and Eq. (\ref{new_d}) imply that $\rho_{11}\cong \rho_{44}$ whenever our large $\bar n$ approximation is valid.  Now we rewrite $C_nC_{n+1}$ as,
\beqa
C_nC_{n+1} &=& \frac{1}{2}\Big\{\cos[gt(\sqrt{n}+\sqrt{n+1})] \nonumber \\
&+&\cos[gt(\sqrt{n+1}-\sqrt{n})]\Big\},
\eeqa
and use the peaked nature of $A_n$ to focus on those terms near to $\bar n$ to introduce the approximation
\beq
\sqrt{n+1}=\sqrt{n}+\frac{1}{2\sqrt{n}},
\label{appr_sqrt}
\eeq
which compresses $C_nC_{n+1}$ to
\beq
C_nC_{n+1}\cong\frac{1}{2}\Big[\cos(2gt\sqrt{n}) + \cos\Big(\frac{gt}{2\sqrt{n}}\Big)\Big].
\eeq
Similarly,
\beqa
S_nS_{n+1}&\cong&\frac{1}{2}\Big[\cos\Big(\frac{gt}{2\sqrt{n}}\Big)-\cos(2gt\sqrt{n})\Big] \nonumber \\
S_nC_{n+1}&\cong&\frac{1}{2}\Big[\sin(2gt\sqrt{n}) - \sin\Big(\frac{gt}{2\sqrt{n}}\Big)\Big] \nonumber \\
S_{n+1}C_{n}&\cong&\frac{1}{2}\Big[\sin(2gt\sqrt{n}) + \sin\Big(\frac{gt}{2\sqrt{n}}\Big)\Big].
\eeqa

With these results we can simplify $\rho_{23}$ further:
\beqa \label{z_sum}
z &\cong& \frac{1}{4}\Big[\Big(\sum_{n}A_n^2 \cos(\frac{gt}{2\sqrt{n}})\Big)^2+\Big(\sum_{n}A_n^2 \sin(\frac{gt}{2\sqrt{n}})\Big)^2 \nonumber \\
&& +\Big(\sum_{n}A_n^2 \cos(2gt\sqrt{n})\Big)^2 \nonumber \\
&&- \Big(\sum_{n}A_n^2 \sin(2gt\sqrt{n})\Big)^2\Big].
\eeqa

Now, using the identities,
\beqa
C_n^2&=&\frac{1+\cos(2gt\sqrt{n})}{2} \nonumber \\
S_n^2&=&\frac{1-\cos(2gt\sqrt{n})}{2},
\eeqa
we can rewrite $\rho_{11}$ and $\rho_{44}$ as,
\beqa \label{a_sum}
\rho_{11} &\cong& \rho_{44} \cong \frac{1}{4}\Big[1-\Big(\sum_{n}A_n^2 \cos(2gt\sqrt{n})\Big)^2 \nonumber \\
&&+ \Big(\sum_{n}A_n^2 \sin(2gt\sqrt{n})\Big)^2\Big].
\eeqa
Then Eqs. (\ref{z_sum}) and  (\ref{a_sum}) lead to:
\beqa\label{z_ad}
\rho_{23}-\sqrt{\rho_{11}\rho_{44}} &\cong& \frac{1}{4} \Big[\Big(\sum_{n}A_n^2 \cos(\frac{gt}{2\sqrt{n}})\Big)^2 \nonumber \\
&&+\Big(\sum_{n}A_n^2 \sin(\frac{gt}{2\sqrt{n}})\Big)^2\nonumber\\
&&+ 2\Big(\sum_{n}A_n^2 \cos(2gt\sqrt{n})\Big)^2 \nonumber \\
&&- 2\Big(\sum_{n}A_n^2 \sin(2gt\sqrt{n})\Big)^2-1\Big].
\eeqa

We can calculate the sums involved here by rewriting them as integrals, treating the integer $n$ as continuous, again relying on the large value of $\bar n$. The first two integrals we need to calculate are,
\beq\label{int1}
I_1=\int_0^{\infty}A_n^2 \cos(\frac{gt}{2\sqrt{n}}) dn, \quad {\rm and}
\eeq
\beq\label{int2}
I_2=\int_0^{\infty}A_n^2 \sin(\frac{gt}{2\sqrt{n}}) dn.
\eeq
We will combine these integrals, $I_1 + iI_2 \equiv I_{12}$ in order to work with the exponential of $igt/2\sqrt{n}$. This, together with Stirling's approximation on $A_n^2$, and the abbreviation $\tau \equiv gt$, leads to
\beq \label{int1-2}
I_{12} \cong \int_0^{\infty} e^{-\alpha^2}\frac{\alpha^{2n}}{\sqrt{2\pi n}} \frac{e^n}{n^{n}} e^{i\tau/2 \sqrt{n}} dn .
\eeq

The saddle point method is appropriate to calculate this integral, and some details are reserved for the Appendix.  The expression for $I_1 + iI_2$ is found to be,
\beq \label{eq.A.35}
I_{12} \cong \exp\Big (-\frac{\tau^2}{32\alpha^4}\Big ) e^{i\tau/2\alpha}.
\eeq

Then helpful cancellations can be identified, and we obtain an
approximate expression for $|\rho_{23}|-\sqrt{\rho_{11}\rho_{44}}$, the entanglement determiner:
\beqa \label{eq.Lambda}
|\rho_{23}|-\sqrt{\rho_{11}\rho_{44}} &\cong& \frac{1}{4}\Big[e^{-g^2t^2/16\bar{n}^2} - 1
\Big]\nonumber\\
& + &\frac{1}{2}\Big[\sum_{n}A_n^2 \cos(2gt\sqrt{n})\Big]^2 \nonumber\\
& - & \frac{1}{2}\Big[\sum_n A_n^2 \sin(2gt\sqrt{n})\Big]^2.
\eeqa

The summations in (\ref{eq.Lambda}) involving $\cos(2gt\sqrt{n})$ and $\sin(2gt\sqrt{n})$ can be combined into a single exponent containing the argument $2igt\sqrt{n}$, which is similar to that in (\ref{int1-2}), but the resulting saddle point analysis is more complicated because now $\sqrt n$ is in the numerator. Details are relegated to the Appendix, which leads to the following working formula:
\beqa \label{C vs T}
&& |\rho_{23}|-\sqrt{\rho_{11}\rho_{44}} \cong  \frac{1}{4}\Big\{\exp\Big (-\frac{\tau^2}{16\alpha^4}\Big) \nonumber \\
&& -1 + e^{-\tau^2/2}\cos(4\alpha\tau)\Big\} \nonumber\\
&&+\sum_{k=1,2,...}\frac{1}{2\pi k}\Big[\exp\Big(-\frac{2(\tau-2\pi k\alpha)^2}{1+\pi^2 k^2}\Big) \nonumber \\
&& \hspace{.25in} \times \cos[4\alpha(\tau-2\pi k\alpha)]
\Big].
\eeqa
In writing this last expression we have used the fact that around $\tau=2\pi k\alpha$ only the term with the corresponding $k$ gives a significant contribution to the sums.  The contribution to $\tau=2\pi k\alpha$ from any other $k^{\prime}$ is proportional to $\exp\Big(-4\pi^2 \alpha^2 (k-k^{\prime})^2/[1+\pi^2 (k^{\prime})^2]\Big)$, so it decays exponentially with the distance from $k$. Thus the final term is the main result, which can be read separately for each value of the step index $k$.

\section{Overview and Implications}

\begin{figure}[t!]
\centering \includegraphics[scale=0.5]{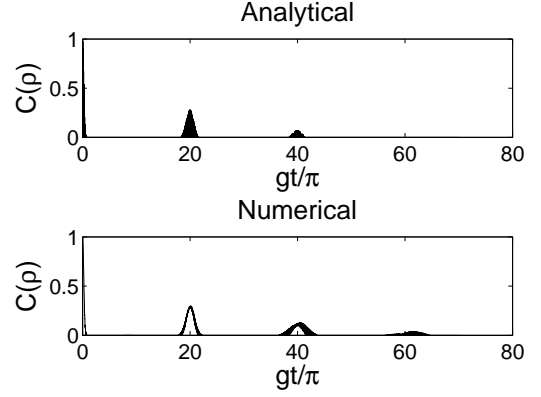}
     \caption{The analytical and numerical results for concurrence of two qubits exposed to two quantum-coherent driving fields with $\alpha = 10$. The evident revivals are predicted reasonably well by the approximate formula (\ref{C vs T}).  Analytical results are for the $X$-type $\rho$ while the numerical ones are for the original $\rho$.}
     \label{revivals 10}
\end{figure}

The $X$-state simplification and Fock-state Ansatz we introduced work together in a surprisingly accurate way. The evaluation of concurrence for $\bar n \gg 1$ here is generically the same as that presented for zero detuning qubit inversion in the original discussion of quantum revivals \citep{Eberly-etal}, and revivals arise in Fig. \ref{revivals 10} because of our focus on strong control and long-time dynamics.  The figure shows both the predictions of the analytic expressions given above, and also the results of a numeric check of the Fock state Ansatz that initiated the analytic calculations.

While entanglement recovery is never complete, the question how complete is important, and it was posed by Lee, et al. \cite{Kim-etalPRL06}. We can display the answer in another analytic formula, giving relative revival envelope heights:
\beq\label{rev_env}
\frac{1}{\pi k} - \frac{1-\exp (-\tau^2/16\alpha^4)}{2}.
\eeq
where $k$ is the revival number.

\begin{figure}[b!]
    \includegraphics[scale=0.5]{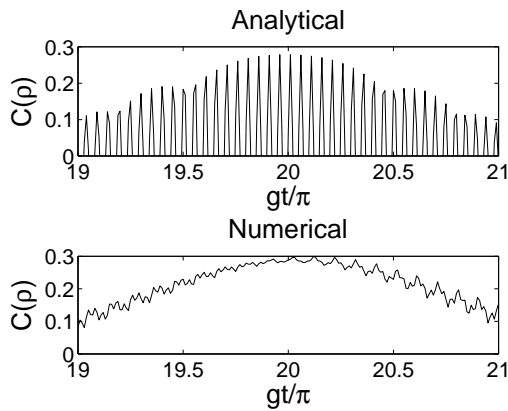}
\caption{Details of the first revival shown in the preceding figure. The deep modulations in the analytic envelope are not present in the numerical envelope, which retains entanglement robustly through the revival event.}
\label{revivals 10detail}
\end{figure}

We also note that while revivals are quite robust, the exact-numeric to approximate-analytic agreements are not perfect, and the differences between them are illuminating. We refer to the absence of Rabi-type behavior during the revivals in the numeric results, which is not predicted analytically. That is, the ESD events (returns of concurrence to zero) that occur within the analytic revival envelope are not present in the numeric envelope. This is shown in the expanded views of the first revival in Fig. \ref{revivals 10detail}.

The absence of Rabi-type behavior in the numeric curves is seen clearly by inspecting a revival envelope in detail, as is done in Fig. \ref{revivals 10detail}.  The analytical results show rapid Rabi-type oscillations with period $\tau = \pi/(2g\sqrt{\bar{n}})$. The analytic formula retains the entanglement death and rebirth episodes that occur on the rapid Rabi-period scale, as were shown in Fig. \ref{eeggESD} and have been discussed in the literature repeatedly for few-photon excitation. By contrast, the numeric results show a smoothed version without rapidly recurring ESD events.  Even the zero revival, i.e., the period that is referred to as the Cummings collapse in the inversion literature, shows no ESD events within it.

Revivals have been demonstrated experimentally for values of $\bar n \ge 3$ (see, e.g., \cite{Winelandgroup}). Our results extend the revival results reported briefly by Paternostro, et al. \cite{Kim-etalPRL06}, which were restricted to shorter times and values of $\bar n$ too small to exhibit the wide revival separations in Fig. \ref{revivals 10}.

Finally, the quasi-periodic modulations evident in the numeric details in Fig. \ref{revivals 10detail} deserve comment. For contrast, we show in Fig. \ref{revivals 5,6detail} below the numeric revival details for two other values of coherent state photon number: $\bar n$ = 25 and 36. In those graphs modulations also appear, but with different main periods, their frequency increasing linearly as $\alpha$ increases, viz., $n$ main modulation periods for $\alpha = n$ over a unit interval in $gt/\pi$. These modulations are an artifact of the assumption that no difference exists between the $\bar n$ values at the two sites being managed. This is an instance where a treatment with more options than presented here (e.g., one with different $\bar n$s) should be expected to promote a desirable feature, the decrease in mid-envelope modulations.

\begin{figure}[t!]
    \includegraphics[scale=0.4]{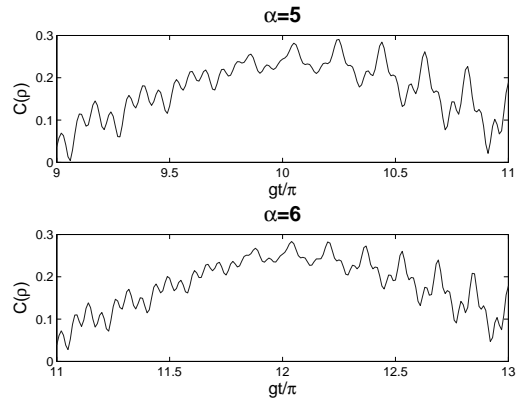}
\caption{Similar to the bottom curve in Fig. \ref{revivals 10detail}, except that here $\alpha$ = 5 and 6, rather than 10.}.\label{revivals 5,6detail}
\end{figure}

\section{Acknowledgements}
We are pleased to thank Prof. Ting Yu for consultation and collaboration in the early stages of this study. Partial financial support was provided by grants from DARPA  HR0011-09-1-0008 and ARO W911NF-09-1-0385.

\section{Appendix}

Here, we sketch the saddle point analysis of the remaining sums in Eqn. (\ref{eq.Lambda}). They can be combined, and converted to an integral: \beq \label{int3+4}
I_{34}  \cong  \int_0^{\infty} \exp{[\alpha^2 f(n)]} dn,
\eeq
where
\beqa \label{fn}
f(n) &=& \frac{1}{\alpha^2} \Big(2n \ln\alpha -\frac{1}{2}\ln(2\pi n) \nonumber \\
& - & n\ln (n) + n + 2i\tau\sqrt{n}\Big) -1.
\eeqa
In the saddle point method, an integral of type $\int_a^b e^{Mf(x)} dx$, where $M$ is a large number and $f(x)$ is a twice-differentiable function, is approximated by,
\beq\label{saddle}
\int_a^b e^{Mf(x)} dx\approx \sqrt{\frac{2\pi}{M|f''(x_0)|}}e^{Mf(x_0)},
\eeq
where $x_0$ is the global maximum of $f(x)$. In our application
Eq. (\ref{int3+4}) can be written in this form if we choose $M=\alpha^2$ and $f(n)$ as given above. For the saddle point maximum of $f(n)$ we need to find the point $n_0$ where $f'(n_0)=0$:
\beq\label{der_f}
f'(n_0) = \frac{1}{\alpha^2} \Big(2\ln\alpha - \frac{1}{2n_0} - \ln (n_0) + i\tau n_0^{-1/2}\Big) = 0.
\eeq
Assuming that $|n_0|$ is large, we can replace
$\ln (n_0) \cong \ln (\alpha^2) + i\tau n_0^{-1/2}$, and then letting $n_0=\rho e^{i\theta}$ gives
\beq
\ln \rho + i\theta = \ln (\alpha^2) + i\tau\rho^{-1/2}e^{-i\theta/2},
\eeq
where equating the real and imaginary parts of both sides gives,
\beqa \label{rho_th}
\ln(\rho)&=&\ln(\alpha^2)+\tau\rho^{-1/2}\sin(\theta/2)\nonumber\\
\theta&=&\tau\rho^{-1/2}\cos(\theta/2).
\eeqa

For $\tau=2\pi k\alpha$, where $k$ is an integer, the equations above become,
\beq
\rho = \alpha^2 \quad {\rm and} \quad \theta = (-1)^k 2\pi k.
\eeq
Now, let $\tau=\tau_0+\Delta\tau$ and $\theta=\theta_0 + \Delta\theta$, where $\tau_0=2\pi k\alpha$ and $\theta_0=2\pi k$.  Assuming that both $\Delta\tau$ and $\Delta\theta$ are small, we can write
\beq
\sin(\theta/2)\cong (-1)^k \frac{\Delta\theta}{2} \quad{\rm and}\quad
\cos(\theta/2)\cong (-1)^k.
\eeq
Then Eq. (\ref{rho_th}) turns into,
\beqa\label{rt}
\rho&\cong&\alpha^2 [1+(-1)^k\pi k\Delta\theta] \nonumber\\
\theta&\cong&\theta_0 -\frac{\pi k\Delta\theta}{2} + (-1)^k\frac{\Delta\tau}{\alpha}.
\eeqa
In order to arrive at the second line we have used,
\beq
\rho^{-1/2}\cong\alpha^{-1}\Big ( 1-(-1)^k\frac{\pi k \Delta\theta}{2}\Big),
\eeq
in the second line of Eq. (\ref{rho_th}) and we have ignored the terms with $\Delta\tau \Delta\theta$.

Now, the last two terms in the second line of Eq. (\ref{rt}) are just $\Delta\theta$, so,
\beq
 -\frac{\pi k\Delta\theta}{2} + (-1)^k\frac{\Delta\tau}{\alpha}=\Delta\theta.
\eeq
Thus, we have,
\beq
\Delta\theta=\frac{(-1)^k}{\alpha(1+\pi^2k^2)}\Delta\tau,
\eeq
Using this in Eq. (\ref{rt}), we obtain
\beqa
\rho&\cong&\alpha^2 [1+\pi k\frac{1}{\alpha(1+\pi^2k^2)}\Delta\tau]\nonumber\\
\theta&=&\theta_0 + (-1)^k\frac{1}{\alpha(1+\pi^2k^2)}\Delta\tau.
\eeqa
Now, by inserting $n_0=\rho e^{i\theta}$ in Eq. (\ref{fn}), we find
\beqa
\alpha^2 f(n_0)&=&-\rho e^{i\theta}\ln(\rho/\alpha^2) - \ln(\rho^{1/2})-\frac{i\theta}{2}\nonumber\\
&&\rho e^{i\theta} (1-i\theta) + 2i\tau \rho^{1/2}e^{i\theta/2} \nonumber\\
&-& \alpha^2 -\ln(\sqrt{2\pi}).
\eeqa

The real part of this equation is given by,
\beqa
Re\{\alpha^2 f(n_0)\} & = & -\rho\cos \theta \ln(\rho/\alpha^2)-\ln(\rho^{1/2}) + \rho\theta\sin\theta \nonumber\\
&&+ \rho\cos\theta-2\tau\rho^{1/2}\sin(\theta/2)\nonumber\\
&&-\alpha^2-\ln(\sqrt{2\pi}).
\eeqa
Using $\theta=\theta_0 + \Delta\theta$ and retaining only the terms upto the second order in $\Delta\theta$,
\beqa
&& Re\{\alpha^2 f(n_0)\} \cong -\rho \Big[1-\frac{(\Delta\theta)^2}{2}\Big] \ln\Big(\frac{\rho}{\alpha^2}\Big) \nonumber \\
&& -\ln(\rho^{1/2}) + \rho\Delta\theta(\theta_0+\Delta\theta)
+\rho\Big[1-\frac{(\Delta\theta)^2}{2}\Big]\nonumber \\
&&-(-1)^k\tau\rho^{1/2}\Delta\theta -\alpha^2-\ln(\sqrt{2\pi}).
\eeqa
Writing $\Delta\theta$ in terms of $\Delta\tau$ and ignoring the terms after second order,
\beqa
&& Re\{\alpha^2 f(n_0)\} \cong -\frac{1}{1+\pi^2 k^2}\Big(\frac{1+2\pi^2 k^2}{2+2\pi^2 k^2}\Big)(\Delta\tau)^2 \nonumber \\
&& -\ln(\rho^{1/2}) -\ln(\rho^{1/2}) -\ln(\alpha^{1/2})-\ln(\sqrt{2\pi}).
\eeqa
The imaginary part of $\alpha^2 f(n_0)$ is,
\beqa
&&Im\{\alpha^2 f(n_0)\} = -\sin\theta\rho\ln(\rho/\alpha^2)-\theta/2\nonumber\\
&&-\rho\theta\cos\theta+\rho\sin\theta+2\tau\rho^{1/2}\cos(\theta/2).
\eeqa
Again, using $\theta=\theta_0+\Delta\theta$ and $\tau=\tau_0+\Delta\tau$, writing $\Delta\theta$ in terms of $\Delta\tau$ and ignoring the terms after second order,
\beqa
&& Im\{\alpha^2 f(n_0)\} \cong (-1)^k\Big\{2\pi k\alpha^2 + \Big[2\alpha-\frac{1}{2\alpha(1+\pi^2k^2)}\Big]\Delta\tau\nonumber\\
&& +\frac{\pi k}{1+\pi^2 k^2}\Big[\frac{3}{2(1+\pi^2 k^2)}-\frac{1}{\alpha^2}\Big](\Delta\tau)^2 \Big\}.
\eeqa
For $k=0$ the $(\Delta\tau)^2$ part of this equation vanishes.  For $k=1,2,...$ this part can be ignored as well.  Thus we are left with,
\beq
Im\{\alpha^2 f(n_0)\}\cong(-1)^k(2\pi k\alpha^2 + 2\alpha\Delta\tau).
\eeq

We also need to calculate
$\sqrt{2\pi/\alpha^2 |f''(n_0)|}$. By using Eq. (\ref{der_f}), we find
\beq
f''(n_0) = \frac{1}{\alpha^2}\Big(\frac{1}{2n_0^2} - \frac{1}{n_0}-\frac{i\tau n_0^{-3/2}}{2}\Big),
\eeq
which has two forms: for $\tau=0$,
\beq
\sqrt{\frac{2\pi}{\alpha^2 |f''(n_0)|}}\cong\sqrt{2\pi\alpha},
\eeq
and for $\tau=2\pi k\alpha$ (k=1,2,...):
\beq
\sqrt{\frac{2\pi}{\alpha^2 |f''(n_0)|}}\cong\sqrt{\frac{2\pi\alpha}{\pi k}}.
\eeq
Then the integral in Eq. (\ref{int3+4}) is given by,
\beq
I_{34} \cong \sqrt{\frac{2\pi}{\alpha^2 |f''(n_0)|}}e^{\alpha^2 f(n_0)}.
\eeq
In order to find the value of this integral we should add the contributions from all $k=0,1,2,...$.  As a result, the integral is,
\beqa
I_{34} &\cong& e^{-\tau^2/2}e^{2i\alpha\tau}\nonumber\\
&+& \sum_{k=1,2,...}\sqrt{\frac{1}{\pi k}}\exp\Big[-\frac{(\tau-2\pi k\alpha)^2}{1+\pi^2 k^2}\Big]\nonumber \\
& \times & \cos[2\alpha(\tau-2\pi k\alpha)].
\eeqa

Now, inserting the values of all four integrals into Eq. (\ref{z_ad}),
\beqa
|\rho_{23}|-\sqrt{\rho_{11}\rho_{44}} &\cong& \frac{1}{4}\Big\{\exp\Big (-\frac{\tau^2}{16\alpha^4}\Big )-1+e^{-\tau^2}\cos(4\alpha\tau)\nonumber\\
&+& \frac{2}{\pi}\sum_{k=1,2,...}\frac{1}{k}\Big[\exp\Big(-\frac{2(\tau-2\pi k\alpha)^2}{1+\pi^2 k^2}\Big) \nonumber \\
& \times & \cos[4\alpha(\tau-2\pi k\alpha)]
\Big]\Big\}.
\eeqa
Writing this last equation we have used the fact that around $\tau=2\pi k\alpha$ only the term with the corresponding $k$ gives a significant contribution to the squares of the sums in Eq. (\ref{z_ad}).  The contribution to $\tau=2\pi k\alpha$ from any other $k^{\prime}$ is proportional to $\exp(-4\pi^2 \alpha^2 (k-k^{\prime})^2/[1+\pi^2 (k^{\prime})^2])$, so it decays exponentially with the distance from the peaks identified with integer $k$.  Thus, while taking the squares in Eq. (\ref{z_ad}) we can ignore the cross-terms.


Now, we turn to the integral given in Eq. (\ref{int1-2}).  We will continue to use the saddle point method.  This time we need to find the maximum of the function:

\beqa \label{fn}
f(n)&=&\frac{1}{\alpha^2} (2n \ln(\alpha) -\frac{1}{2}\ln(2\pi n)\nonumber\\
&&-n\ln (n) + n + \frac{i\tau}{2 \sqrt{n}})-1,
\eeqa
For this, we need to find the point $n_0$ where $f'(n_0)=0$:
\beq
f'(n_0)=\frac{1}{\alpha^2} (\ln (\alpha^2) - \frac{1}{2n_0} - \ln (n_0) - \frac{i\tau}{4} n_0^{-3/2})=0,
\eeq
Again assuming that $|n_0|>>1$,
\beq
\ln (n_0) \cong \ln (\alpha^2) - \frac{i\tau}{4} n_0^{-3/2}.
\eeq
Letting
\beq
n_0=\rho e^{i\theta},
\eeq
then,
\beq
\ln(\rho) + i\theta = \ln(\alpha^2) - \frac{i\tau}{4} \rho^{-3/2}[\cos(\frac{3\theta}{2})-i\sin(\frac{3\theta}{2})],
\eeq
and by matching the real and imaginary parts of both sides, we find two coupled transcendental equations:
\beqa
\ln(\rho) &=& \ln(\alpha^2) - \frac{\tau}{4} \rho^{-3/2}\sin(\frac{3\theta}{2}), \nonumber \\
\theta &=& - \frac{\tau}{4} \rho^{-3/2}\cos(\frac{3\theta}{2}).
\eeqa
We are going to retain only the terms up to second order in $\tau \rho^{-3/2}$,
\beqa
\theta \cong - \frac{\tau}{4\alpha^3}\quad {\rm and}\quad
\rho \cong \alpha^2 (1+\frac{3\theta^2}{2}).
\eeqa
With this restriction, and after inserting $n=\rho e^{i\theta}$ into Eq. (\ref{fn}), we find
\beqa
\alpha^2f(n_0)&\cong& -ln(\alpha^2)-\ln(\alpha) -\ln(\sqrt{2\pi})\nonumber\\
&&-\frac{3}{4}\theta^2 - \frac{i\theta}{2}+\rho e^{i\theta}(1-\frac{3\theta^2}{2} - i\theta)\nonumber\\
&&+ \frac{i\tau}{2}\rho^{-1/2}e^{-i\theta/2} - \alpha^2.
\eeqa
Writing $\rho$ in terms of $\theta$ and retaining only the terms up to $\theta^2$,
\beqa
\alpha^2f(n_0)&\cong&-ln(\alpha^2)-\ln(\alpha)-\ln(\sqrt{2\pi})\nonumber\\
&&-\frac{3\theta^2}{4}+\frac{\alpha^2\theta^2}{2}+\frac{\tau\theta}{4\alpha}\nonumber\\
&&+i\Big ( \frac{\tau}{2\alpha}-\frac{\theta}{2}-\frac{7\tau\theta^2}{16\alpha}\Big ),
\eeqa
so that by inserting $-\tau/4\alpha^3$ for $\theta$, we obtain
\beqa\label{mfn}
\alpha^2f(n_0)&\cong&-ln(\alpha^2)-\ln(\alpha)-\ln(\sqrt{2\pi})\nonumber\\
&&-\frac{\tau^2}{32\alpha^4}-\frac{3\tau^2}{16\alpha^6}\nonumber\\
&&+i\Big (\frac{\tau}{2\alpha}
-\frac{\tau}{8\alpha^3}-\frac{7\tau^3}{256\alpha^7}\Big).
\eeqa
In order to find the final form of Eq. (\ref{saddle}) we need to calculate $|f''(n_0)|$ as well.
\beq
f''(n_0)=\frac{1}{\alpha^2}\Big ( \frac{1}{2n_0^2}-\frac{1}{n_0}+\frac{3i\tau}{8}n_0^{-5/2} \Big ).
\eeq
Retaining the terms up to $\alpha^{-2}$ in the parentheses, we have
$|f''(n_0)|\cong \frac{1}{\alpha^4}$. Then we use Eq. (\ref{mfn}) and  Eq. (\ref{saddle}) and insert $\alpha^2$ for $M$, to get
\beqa
\sqrt{\frac{2\pi}{\alpha^2|f''(n_0)|}}e^{\alpha^2f(n_0)}&\cong& \exp \Big (-\frac{\tau^2}{32\alpha^4}-\frac{3\tau^2}{16\alpha^6} \Big )\nonumber\\
&&\exp \Big (\frac{i\tau}{2\alpha}
-\frac{i\tau}{8\alpha^3}-\frac{7i\tau^3}{256\alpha^7}\Big).
\eeqa
For $\tau\sim\alpha^2$ we can retain the first terms in the parentheses and ignore the rest since $\alpha^2 >> 1$.  Thus, we finally find
\beq
I_{12} \cong \exp\Big (-\frac{\tau^2}{32\alpha^4}\Big ) e^{i\tau/2\alpha}.
\eeq



\begin{thebibliography}{99}

\bibitem{Schr} E. Schr\"odinger, {\it Naturwiss}.  {\bf 23},  807
(1935). See also the closely related paper, A. Einstein, B. Podolsky and N. Rosen, \pr {46}, 777 (1935).

\bibitem{Jaynes-Cummings} E.T. Jaynes and F.W. Cummings, Proc. IEEE {\bf 51}, 89 (1963).

\bibitem{Kim-etalJOM02} W. Son, M.S. Kim, J. Lee and D. Ahn, \jmo{49}, 1739 (2002).

\bibitem{Yu-EberlySci09} T. Yu and J.H. Eberly, \sci{323}, 598 (2009).

\bibitem{Krause-Cirac} B. Kraus and J.I. Cirac, \prl{92}, 013602 (2004).

\bibitem{Paternostro-etalPRL04} M. Paternostro, W. Son, M.S. Kim, \prl{92}, 197901 (2004).

\bibitem{Paternostro-etalPRA04} M. Paternostro, W. Son, M.S. Kim, G. Falci and G.M. Palma, \pra{70}, 022320 (2004).

\bibitem{Kim-etalPRL06} J. Lee, M. Paternostro, M.S. Kim and S. Bose, \prl{96}, 080501 (2006).

\bibitem{Zhou-Wang} L. Zhou and G. Yang, \jpb{39}, 5143-5150 (2006).

\bibitem{Rendell-Rajagopal} R. W. Rendell and A. K. Rajagopal, \pra{67}, 062110 (2003).

\bibitem{Rempegroup} G. Rempe, H. Walther and N. Klein, \prl{58}, 353 (1987).

\bibitem{Harochegroup}  M. Brune \emph{et al.}, \prl{76}, 1800 (1996).

\bibitem{Winelandgroup} D. M. Meekhof, C. Monroe, B. E. King, W. M. Itano and D. J. Wineland, Phys.  Rev.  Lett.  \textbf{76}, 1796 (1996).

\bibitem{Kimblegroup} A. Boca \emph{et al.}, Phys.  Rev.  Lett \textbf{93}, 233603 (2004).

\bibitem{Banacloche} J. Gea-Banacloche, \prl{65}, 3385 (1990).

\bibitem{Phoenix-Knight} S.J.D. Phoenix and P.L. Knight, \pra {44}, 6023 (1991).

\bibitem{Wootters} W.K. Wootters, \prl{80}, 2245 (1998).

\bibitem{Yu-EberlyQIC07} T. Yu and J.H. Eberly, \qic{7}, 459 (2007).

\bibitem{Yu-EberlyJMO07} T. Yu and J.H. Eberly, \jmo{54}, 2289 (2007).

\bibitem{Yonac-etal07} M. Y\"ona\c{c}, et al., \jpb{40}, S45 (2007). See also M. Y\"ona\c{c}, et al., \jpb{39}, S621 (2006).

\bibitem{Sainz-Bjork} I. Sainz and G. Bj\"ork, \pra{77}, 052307 (2008).

\bibitem{Chan-etal} S. Chan, M. D. Reid and Z. Ficek, \jpb{42}, 065507 (2009).

\bibitem{JJSM} J.J. Sanchez-Mondragon, N.B. Narozhny and J.H. Eberly, \prl{51}, 550 (1983).

\bibitem{Agarwal84} G.S. Agarwal, \prl{53}, 1732 (1984).

\bibitem{YonacThesis} M. Y\"ona\c{c}, Ph.D. Thesis, Department of Physics and Astronomy, University of Rochester (2009).

\bibitem{Eberly-etal} J.H. Eberly, N.B. Narozhny and J.J. Sanchez-Mondragon, \prl {44}, 1323 (1980), and  N.B. Narozhny, J.J. Sanchez-Mondragon and J.H. Eberly, \pra {23}, 236 (1981).


\end{thebibliography}
\end{document}